\documentclass[twoside]{ilcws08}
\usepackage[latin1]{inputenc}
\usepackage[dvips]{graphicx,epsfig,color}
\usepackage{wrapfig,rotating}
\usepackage{amssymb,amsmath,array}

\begin{document}
\title{Precision Measurements of the model parameters 
in the Littlest Higgs model with T-parity} 
\author{Eri Asakawa$^1$,
        Masaki Asano$^2$,
        Keisuke Fujii$^3$,
        Tomonori Kusano$^4$, \\
        Shigeki Matsumoto$^5$,
        Rei Sasaki$^4$,
        Yosuke Takubo$^4$
    and
        Hitoshi Yamamoto$^4$
\vspace{.3cm}\\
1- Institute of Physics, Meiji Gakuin University, Yokohama, Japan \\
2- Institute for Cosmic Ray Research (ICRR), University of Tokyo, Kashiwa, Japan \\
3- High Energy Accelerator Research Organization (KEK), Tsukuba, Japan \\
4- Department of Physics, Tohoku University, Sendai, Japan \\
5- Department of Physics, University of Toyama, Toyama, Japan \\
}

\maketitle

\begin{abstract}
 We investigate a possibility of precision measurements for parameters of 
 the Littlest Higgs model with T-parity at the International Linear 
 Collider (ILC). The model predicts new gauge bosons which masses strongly 
 depend on the vacuum expectation value that breaks a global symmetry of 
 the model. 
%
 Through Monte Carlo simulations of production processes of new gauge bosons,
 we show that these masses can be determined very accurately at the ILC 
 for a representative parameter point of the model. From the simulation 
 result, we also discuss the determination of other model parameters at the 
 ILC.
\end{abstract}

\section{Introduction}
The Little Higgs model \cite{Arkani-Hamed:2001nc, Arkani-Hamed:2002qy} has 
been proposed for solving the little hierarchy problem. 
In this scenario, the Higgs boson is regarded as a pseudo Nambu-Goldstone (NG) 
boson associated with a global symmetry at some higher scale. Though the 
symmetry is not exact, its breaking is specially arranged to cancel 
quadratically divergent corrections to the Higgs mass term at 1-loop level. 
This is called the Little Higgs mechanism. 
As a result, the scale of new physics can be as high as 10 TeV without a 
fine-tuning on the Higgs mass term.  Due to the symmetry, the scenario 
necessitates the introduction of new particles. 
In addition, the implementation of the $Z_2$ symmetry called T-parity to 
the model has been proposed in order to avoid electroweak precision 
measurements \cite{Cheng:2003ju}.
In this study, we focus on the Littlest Higgs model with T-parity as a 
simple and typical example of models implementing both the Little Higgs 
mechanism and T-parity. 

In order to test the Little Higgs model, precise determinations of 
properties of Little Higgs partners are mandatory, because these particles 
are directly related to the cancellation of quadratically divergent 
corrections to the Higgs mass term. In particular, measurements of heavy 
gauge boson masses are quite important. Since heavy gauge bosons acquire 
mass terms through the breaking of the global symmetry, 
precise measurements of their masses allow us to determine the most 
important parameter of the model, namely the vacuum expectation value (VEV) 
of the breaking. Furthermore, because the heavy photon is a candidate for 
dark matter \cite{Hubisz:2004ft, Asano:2006nr}, the determination of its 
property gives a great impact not only on particle physics but also on 
astrophysics and cosmology. 
However, it is difficult to determine the properties of heavy gauge bosons 
at the Large Hadron Collider, because they have no color charge 
\cite{Cao:2007pv}. 

On the other hand, the ILC will provide an 
ideal environment to measure the properties of heavy gauge bosons. 
We study the sensitivity of the measurements to the Little Higgs parameters 
at the ILC based on a realistic Monte Carlo simulation \cite{Asakawa:2009qb}. 
We have used MadGraph \cite{madgraph} and Physsim \cite{physsim} to 
generate signal and Standard Model (SM) events, respectively. 
In this study, we have also used PYTHIA6.4 \cite{pythia}, 
TAUOLA \cite{tauola} and JSFQuickSimulator which implements the GLD geometry 
and other detector-performance related parameters \cite{glddod}. 

%

%


\section{Model}


The Littlest Higgs model with T-parity is based on a non-linear sigma model 
describing an SU(5)/SO(5) symmetry breaking with a VEV, 
$f \sim {\cal O}(1)$ TeV. An [SU(2)$\times$U(1)]$^2$ subgroup in the SU(5) 
is gauged, which is broken down to the SM gauge group 
SU(2)$_L\times$U(1)$_Y$. Due to the presence of the gauge and Yukawa 
interactions, the SU(5) global symmetry is not exact. The SM doublet and 
triplet Higgs bosons ($H$ and $\Phi$) arise as pseudo NG bosons in 
the model. 
The triplet Higgs boson is T-odd, while the SM Higgs is T-even.

This model contains gauge fields of the gauged [SU(2)$\times$U(1)]$^2$ 
symmetry; The linear combinations $W^a = (W^a_1 + W^a_2)/\sqrt{2}$ and 
$B = (B_1 + B_2)/\sqrt{2}$ correspond 
to the SM gauge bosons for the SU(2)$_L$ and U(1)$_Y$ symmetries. The other 
linear combinations $W^a_{\mathrm{H}} = (W^a_1 - W^a_2)/\sqrt{2}$ and 
$B_{\mathrm{H}} = (B_1 - B_2)/\sqrt{2}$ are additional gauge bosons called 
heavy gauge bosons, which acquire masses of ${\cal O}(f)$ through the 
SU(5)/SO(5) symmetry breaking. After the electroweak symmetry breaking, 
the neutral components of $W^a_{\mathrm{H}}$ and $B_{\mathrm{H}}$ are 
mixed with each other and form mass eigenstates $A_{\mathrm H}$ and 
$Z_{\mathrm H}$. 
The heavy gauge bosons 
($A_{\mathrm H}$, $Z_{\mathrm H}$, and $W_{\mathrm H}$) behave as T-odd 
particles, while SM gauge bosons are T-even.


To implement T-parity, two SU(2) doublets $l^{(1)}$ and $l^{(2)}$ are 
introduced for each SM lepton. The quantum numbers of $l^{(1)}$ and 
$l^{(2)}$ under the gauged [SU(2)$\times$U(1)]$^2$ symmetry are 
 $({\bf 2}, -3/10; {\bf 1}, -1/5)$ and
 $({\bf 1}, -1/5; {\bf 2}, -3/10)$, respectively.  
The linear combination $l_{SM} = (l^{(1)} - l^{(2)})/\sqrt{2}$ gives the 
left-handed SM lepton. 
On the other hand, another linear combination 
$l_{\mathrm{H}} = (l^{(1)} + l^{(2)})/\sqrt{2}$ is vector-like T-odd partner 
which acquires the mass of ${\cal O}(f)$. 
The masses depend on the $\kappa_l$:
$ m_{e_{\mathrm{H}}} = \sqrt{2} \kappa_l f, 
  m_{\nu_{\mathrm{H}}} = (1/2)( \sqrt{2}+\sqrt{1+c_f} ) \kappa_l f
                    \simeq \sqrt{2} \kappa_l f$. 
In addition, new particles are also introduced in quark sector. (For details, 
see Ref. \cite{littlest_review}.)


%

\section{Simulation study}
The representative point used in our simulation study is 
$(f, m_h, \lambda_2, \kappa_l)$ $=$ (580 GeV, 134 GeV, 1.5, 0.5) where 
 (    $m_{A_{\mathrm{H}}}$, $m_{W_{\mathrm{H}}}$, 
      $m_{Z_{\mathrm{H}}}$, $m_{\Phi}$            )
$=$ (81.9 GeV, 368 GeV, 369 GeV, 440 GeV) and $\lambda_2$ is an additional 
Yukawa coupling in the top sector. The model parameter satisfies 
not only the current electroweak precision data but also the WMAP 
observation \cite{Komatsu:2008hk}. Furthermore, no fine-tuning is needed 
at the sample point to keep the Higgs mass on the electroweak scale 
\cite{Hubisz:2005tx,Matsumoto:2008fq}. 
\begin{figure}[t]
 \begin{center}
  \scalebox{0.55}{\includegraphics*{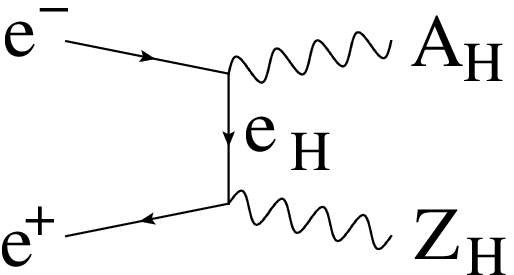}}
  \qquad\qquad
  \scalebox{0.55}{\includegraphics*{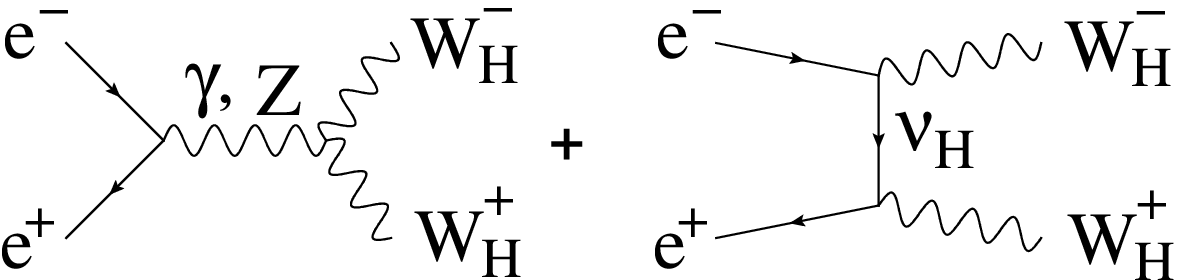}}
\caption{Diagrams for signal processes; 
         $e^+ e^- \rightarrow A_{\rm H} Z_{\rm H}$ and 
        $e^+ e^- \rightarrow W_{\rm H}^+ W_{\rm H}^-$. 
         }\label{Fig:diagram}
 \end{center}
\end{figure}


In the model, 
there are four processes whose final states consist of two heavy gauge bosons: $e^+e^- \rightarrow$ $A_{\mathrm{H}}A_{\mathrm{H}}$, $A_{\mathrm{H}}Z_{\mathrm{H}}$, $Z_{\mathrm{H}}Z_{\mathrm{H}}$, and $W_{\mathrm{H}}^+ W_{\mathrm{H}}^-$. The first process is undetectable. At the representative point, the largest cross section is expected for the fourth process, which is open at $\sqrt{s} > 1$ TeV. On the other hand, because $m_{A_{\mathrm{H}}} + m_{Z_{\mathrm{H}}}$ is less than 500 GeV, the second process is important already at the $\sqrt{s} = 500$ GeV. We, hence, concentrate on $e^+e^- \rightarrow A_{\mathrm{H}}Z_{\mathrm{H}}$ at $\sqrt{s} = 500$ GeV and $e^+e^- \rightarrow W_{\mathrm{H}}^+W_{\mathrm{H}}^-$ at $\sqrt{s} = 1$ TeV. 
Feynman diagrams for the signal processes are shown in Fig. \ref{Fig:diagram}. 

\begin{figure}
 \begin{center}
  \includegraphics[width=10cm]{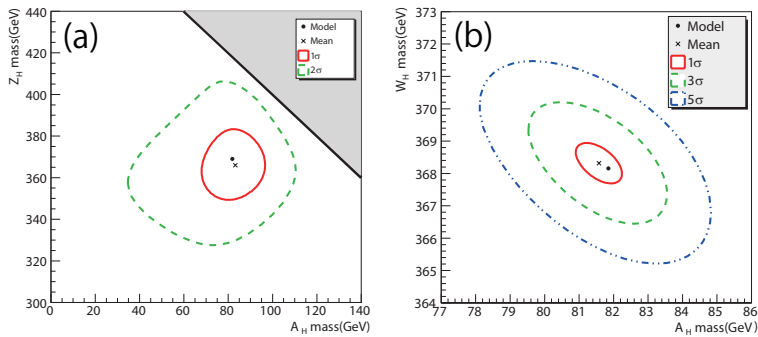}
 \end{center}
 \caption{Probability contours corresponding to (a) 1- and 2-$\sigma$ deviations from the best fit point in the $A_{\mathrm H}$ and $Z_{\mathrm H}$ mass plane, and (b) 1-, 3-, and 5-$\sigma$ deviations in the $A_{\mathrm H}$ and $W_{\mathrm H}$ mass plane. The shaded area in (a) shows the unphysical region of $m_{A_{\mathrm{H}}} + m_{Z_{\mathrm{H}}} > 500$ GeV.}
 \label{fig:cntr}
\end{figure}


For the $A_{\mathrm H} Z_{\mathrm H}$ production at $\sqrt{s} = 500$ GeV
with an integrated luminosity of 500 fb$^{-1}$,
we define $A_{\mathrm H} Z_{\mathrm H} 
           \rightarrow A_{\mathrm H} A_{\mathrm H} h
           \rightarrow A_{\mathrm H} A_{\mathrm H} bb$ 
as our signal event. The $A_{\mathrm H}$ and  $Z_{\mathrm H}$ boson 
masses can be estimated from the edges of the distribution of the 
reconstructed Higgs boson energies. 
The endpoints have been 
estimated by fitting the distribution with a line shape determined by a 
high statistics signal sample. The fit resulted in $m_{A_{\mathrm{H}}}$ 
and $m_{Z_{\mathrm{H}}}$ being $83.2 \pm 13.3$ GeV and $366.0 \pm 16.0$ 
GeV, respectively. 

For the $W_{\mathrm H} W_{\mathrm H}$ production at $\sqrt{s} = 1$ TeV
with an integrated luminosity of 500 fb$^{-1}$,
we have used 4-jet final states, 
$W_{\mathrm{H}}^+W_{\mathrm{H}}^- 
 \rightarrow A_{\mathrm{H}} A_{\mathrm{H}} W^+ W^-
 \rightarrow A_{\mathrm{H}} A_{\mathrm{H}} qqqq$.
The masses of $A_{\mathrm H}$ and $W_{\mathrm H}$ bosons can be 
determined from the edges of the $W$ energy distribution. The fitted masses of $A_{\mathrm{H}}$ and 
$W_{\mathrm{H}}$ bosons are $81.58 \pm 0.67$ GeV and $368.3 \pm 0.63$ 
GeV, respectively. 
Using the process, it is also possible to confirm that the spin of 
$W_{\mathrm H}^{\pm}$ is consistent with one and the polarization of 
$W^{\pm}$ from the $W_{\mathrm H}^{\pm}$ decay is dominantly longitudinal. 
Furthermore, the gauge charges of the $W_{\mathrm H}$ boson could be also 
measured using a polarized electron beam.

Figure \ref{fig:cntr} shows the probability 
contours for the masses of $A_{\mathrm{H}}$ and $W_{\mathrm{H}}$ at 
$1$ TeV together with that of $A_{\mathrm H}$ and $Z_{\mathrm H}$ at 500 GeV. The mass 
resolution improves dramatically at $\sqrt{s} = 1 $ TeV, compared to 
that at $\sqrt{s} = 500$ GeV.

\section{Conclusion}

The Littlest Higgs Model with T-parity is one of the attractive candidates 
for physics beyond the SM. 
We have shown that the masses of the heavy gauge bosons
can be determined very accurately at the ILC.  
It is important to notice that these masses are obtained in a 
model-independent way, so that it is possible to test the Little Higgs 
model by comparing them with the theoretical predictions. 
Furthermore, since the masses of the heavy gauge bosons are determined 
by the VEV $f$, it is possible to accurately determine $f$. 
From the results obtained in our simulation study, it turns out that 
the VEV $f$ can be determined to accuracies of 4.3\% at $\sqrt{s}=$ 
500 GeV and 0.1\% at $\sqrt{s}=$ 1 TeV. 
Another Little Higgs parameter $\kappa_l$
could also be estimated from production cross sections for the heavy 
gauge bosons, because the cross sections depend on the masses of heavy 
leptons. 
At the ILC with $\sqrt{s}=$ 500 GeV and 1 TeV, $\kappa_l$ 
could be obtained within 9.5\% and 0.8\% accuracies, respectively.

Finally, 
We have also found that the thermal abundance of dark matter relics can be 
determined to 10\% and 1\% levels at $\sqrt{s}=$ 500 GeV and $\sqrt{s}=$ 1 TeV, 
respectively. These accuracies are comparable to those of current and future 
cosmological observations such as the PLANCK satellite \cite{Planck:2006uk}, 
implying that the ILC experiment will play an essential role to understand the 
thermal history of our universe.


\section{Acknowledgments}
The authors would like to thank all the members of the ILC physics subgroup
\cite{softg} for useful discussions. 
This study is supported in part by the Creative Scientific Research Grant
No. 18GS0202 of the Japan Society for Promotion of Science.


\begin{footnotesize}



%

\end{footnotesize}


\end{document}